%% file: main.tex
\documentclass[conference]{IEEEtran}
\IEEEoverridecommandlockouts
\usepackage{cite}
\usepackage[hyphens]{url}
\usepackage{hyperref}
\usepackage[hyphenbreaks]{breakurl}
\usepackage{subcaption}
\usepackage{amsmath,amssymb,amsfonts}
\usepackage{algorithmic}
\usepackage{graphicx}
\usepackage{textcomp}
\usepackage{xcolor}
\usepackage{array}
\usepackage{booktabs}
\usepackage{multirow}

\newcommand{\blue}[1]{\textcolor{blue}{#1}}
\newcommand{\red}[1]{\textcolor{black}{#1}}

\begin{document}

\title{An Investigation of Large Language Models for Real-World Hate Speech Detection}

\author{
\IEEEauthorblockN{Keyan Guo\IEEEauthorrefmark{1},
Alexander Hu\IEEEauthorrefmark{3}\IEEEauthorrefmark{4}, 
Jaden Mu\IEEEauthorrefmark{5}\IEEEauthorrefmark{4},
Ziheng Shi\IEEEauthorrefmark{6}\IEEEauthorrefmark{4},
Ziming Zhao\IEEEauthorrefmark{1},
Nishant Vishwamitra \IEEEauthorrefmark{2},
Hongxin Hu\IEEEauthorrefmark{1}
}
\IEEEauthorblockA{
\IEEEauthorrefmark{1}University at Buffalo,
\IEEEauthorrefmark{2}University of Texas at San Antonio, 
\IEEEauthorrefmark{3}Union County Magnet High School,\\
\IEEEauthorrefmark{5}East Chapel Hill High School,
\IEEEauthorrefmark{6}Minhang Corsspoint Academy\\
\{keyanguo, zimingzh, hongxinh\}@buffalo.edu,
\{chujie.hu, jaden.mu\}@gmail.com,
ziheng\_shi@qq.com,\\
nishant.vishwamitra@utsa.edu
}}

\maketitle

\begingroup\renewcommand\thefootnote{\textsection}
\footnotetext{
Work done during internship at University at Buffalo.}
\begin{abstract}

Hate speech has emerged as a major problem plaguing our social spaces today.
%
While there have been significant efforts to address this problem, existing methods are still significantly limited in effectively detecting hate speech online.
A major limitation of existing methods is that hate speech detection is a highly contextual problem, and these methods cannot fully capture the context of hate speech to make accurate predictions.
Recently, large language models (LLMs) have demonstrated state-of-the-art performance in several natural language tasks.
%
LLMs have undergone extensive training using vast amounts of natural language data, enabling them to grasp intricate contextual details. Hence, they could be used as knowledge bases for context-aware hate speech detection. However, a fundamental problem with using LLMs to detect hate speech is that there are no studies on effectively prompting LLMs for context-aware hate speech detection.
In this study, we conduct a large-scale study of hate speech detection, employing five established hate speech datasets. We discover that LLMs not only match but often surpass the performance of current benchmark machine learning models in identifying hate speech.
By proposing four diverse prompting strategies that optimize the use of LLMs in detecting hate speech.
Our study reveals that a meticulously crafted reasoning prompt can effectively capture the context of hate speech by fully utilizing the knowledge base in LLMs, significantly outperforming existing techniques. 
Furthermore, although LLMs can provide a rich knowledge base for the contextual detection of hate speech, suitable prompting strategies play a crucial role in effectively leveraging this knowledge base for efficient detection.

\end{abstract}

\begin{IEEEkeywords}
hate speech, large language model, prompt engineering, few-shot learning
\end{IEEEkeywords}

\section{Introduction}

The emerging threat of online hate is increasingly disrupting the daily experiences of internet users, serving as a grim reminder of the darker aspects of digital interactions~\cite{thomas2021sok}. 
Hate speech stands out as a particularly damaging form of online abuse, causing emotional harm, facilitating social discrimination, and even inciting violent behavior~\cite{10.1371/journal.pone.0221152}. 
Despite sustained efforts by social media platforms to enhance their policies and improve detection mechanisms~\cite{twitter_hate_speech_policy, twitter_hate_speech_AI, facebook_hate_speech_AI}, they continue to struggle with mitigating hate speech effectively, owing to its complex and ever-changing nature. 
Consequently, the urgent development of more effective hate speech detectors is necessary and pressing.
Encouragingly, the rapid strides in artificial intelligence (AI) and machine learning (ML) technologies have spawned many studies~\cite{elsherief2018hate,nobata2016abusive} that have achieved promising results in this domain.
However, state-of-the-art AI models predominantly rely on supervised learning techniques~\cite{schmidt-wiegand-2017-survey},  
which are only capable of performing simple binary predictions of hate speech.
A critical challenge in AI-based hate speech detection is that it is a highly contextual problem, and existing supervised learning-based detection methods cannot fully capture this context to make accurate predictions. As a result, there is a need for detection methods that can detect hate speech by fully capturing its context.\looseness=-1

Recently, the rise of large language models (LLMs) like ChatGPT has been a significant development in the field of Natural Language Processing (NLP)\cite{openai2022gptblog}. LLMs have exhibited remarkable proficiency across various NLP tasks, from text classification to sentiment analysis\cite{bang2023multitask, guo2023close}. 
Their efficacy can be attributed to the extensive training data they are trained on, which allows for these models to learn a vast \textit{knowledge base} of human language~\cite{zhao2023survey}. 
The knowledge base acquired by LLMs provides advanced contextual understanding, which allows performing complex NLP tasks. 
However, a critical question arises: ``How can this knowledge base be effectively leveraged to perform contextual detection of hate speech?''
Recently, prompt-based strategies have been found to effectively guide LLMs in leveraging the context of the specific task at hand~\cite{10.1145/3560815}. 
A prompt is essentially a query or statement designed to instruct the model on what is being asked. The effectiveness of an LLM can be significantly improved with carefully crafted prompts, underscoring the importance of prompting techniques in leveraging the context of these models.
However, the effectiveness of different prompting strategies on leveraging the knowledge base of LLMs in the contextual detection of hate speech has not been systematically explored.

In this work, we conduct the first large-scale investigation into the contextual detection of hate speech using LLMs. 
The major goal of our work is to explore the prompting strategies that are most effective in leveraging the knowledge base in LLMs to contextually detect hate speech.
To this end, we base our analyses on five real-world hate speech datasets, 
and rigorously explore the effectiveness of four novel prompting strategies on their ability to leverage the knowledge base in LLMs for contextually identifying hate speech. 
This approach allows us not only to assess the general aptitude of LLMs in this vital area but also to understand how variations in prompt design can significantly influence detection performance.

Based on our extensive experiments, several key insights emerged: 
First, LLMs not only match but, in certain instances, surpass the performance of state-of-the-art hate speech detection models, specifically on the HateXplain datasets, showcasing a substantial 7.9\% to 24.2\% improvement of F1 Score in flagging hate speech.
Second, the effectiveness of LLMs in identifying hate speech is highly contingent upon the design of the prompt. A prompt that is meticulously constructed, incorporating chain-of-thought reasoning, can significantly enhance the model’s ability to identify hate speech, as evidenced by an impressive F1 Score of 0.87.
Third, while LLMs exhibit strong capabilities in multilingual processing, such as multilingual machine translation, they underperform in hate speech detection in non-English text, pointing to an area requiring further investigation.

\section{Background and Related Work}

\subsection{Hate Speech Detection}
In recent times, online hate, especially hate speech, has emerged as a critical threat~\cite{thomas2021sok} that has been the focus of both governments~\cite{doj_hate_crime_laws} and institutions~\cite{twitter_hate_speech_policy,fb_hate_speech_def}. 
Recent surveys reveal that a sizable portion of the population has experienced online harassment, both in the United States and globally~\cite{duggan2017online, microsoft_civility}. Initially, the primary defense against this growing issue relied on human moderators~\cite{fb_community_standards_enforcement_report, healthier_twitter}. However, this approach faces challenges in scalability and ethical considerations~\cite{content_moderators}.
Consequently, artificial intelligence and machine learning (AI/ML) have emerged as promising technologies for mitigating online hate~\cite{ elsherief2018hate, nobata2016abusive}. 
While there is ongoing research on hate speech detection, a notable challenge lies in the varying definitions of hate speech among different groups. The lack of a unified definition, particularly when considering various types of hate speech, complicates the problem significantly~\cite{10.1371/journal.pone.0221152}\cite{thomas2021sok}.
These AI/ML detectors, which heavily rely on supervised learning techniques, face significant limitations due to these challenges. Their effectiveness is strongly influenced by the labeling strategies of their training sets, leading to inconsistent performance when confronted with different forms of hate speech. 
Given these limitations, there is a pressing need for more robust solutions to effectively address the complex and evolving issue of hate speech.

\subsection{LLMs and Prompts-based Hate Speech Detection}

Recently, there has been a growing interest in the use of LLMs for hate speech detection.
A notable study evaluated GPT-3's ability to identify sexist and racist text, finding that the model could reach an accuracy of approximately 85\% through few-shot learning~\cite{chiu2022detecting}.
However, this study was constrained by its focus on specific types of hate speech and limited test data, as well as an absence of systematic analysis. 
Li et al.~\cite{li2023hot} initiative assessed ChatGPT's ability to classify harmful content, such as hateful, offensive, and toxic speech, and found results comparable to expert annotations from Amazon Mechanical Turk. 
Another study by He et al.~\cite{he2023prompt} explored the application of LLMs and prompt learning in addressing the issue of toxic content. Utilizing models like GPT-3 and T5, their approach outperformed the best baseline, achieving an F1 Score of 0.643 compared to 0.640.
While these studies make valuable contributions, our research aims to offer a more comprehensive understanding of LLMs' proficiency in hate speech detection. 
We further extend the current discourse by rigorously evaluating various prompting strategies, such as few-shot learning and chain-of-thought reasoning, which have been examined in other contexts but not specifically for hate speech detection~\cite{10.1145/3560815, zhong2023chatgpt}. 
Huang et al.~\cite{Huang_2023} utilized chain-of-thought reasoning to interpret implicit hate speech and reported promising findings. 
Unlike this research, however, our work is uniquely centered on the identification of explicit hate speech, addressing a critical gap in the realm of online safety.



\input{data}

\input{prompt}

\input{evaluation}



\section{Conclusion and Future Work}

In conclusion, this paper presents the first extensive study on effectively leveraging LLMs for the detection of hate speech. 
Our work fills an important gap in the literature, providing a detailed examination of various prompting strategies and their effectiveness in hate speech detection. 
To this end, we conducted a large-scale study based on five hate speech datasets and four prompting strategies and studied the effectiveness of these strategies in leveraging the knowledge base in LLMs for contextual detection of hate speech.
Our study reveals that the prompting strategies play a key role in effectively leveraging LLMs to detect hate speech.
The insights gained from this study are crucial stepping stones toward improving the reliability and effectiveness of automated hate speech detection systems, especially as LLMs continue to be more pervasive in language-based applications.

In the future, we plan to continue our research in this critical domain. One primary focus will be to enhance the capabilities of LLMs in multilingual settings, particularly for hate speech detection. In addition, given the evolving nature of online hate, as evidenced by current studies that highlight the rise of hateful memes as a significant issue~\cite{10179315, kiela2021hateful, osti_10399964}, we aim to leverage LLMs in detecting such multimodal content. Through these future endeavors, we hope to contribute further to the development of more robust and reliable automated systems for combating online hate in its various forms. 

\section*{Acknowledgements}
This material is based upon work supported in part by the National Science Foundation (NSF) under Grant No. 2129164, 2114982, 2228617, 2120369, 2237238, and 2245983.




\bibliographystyle{IEEEtran}
\bibliography{bibtexes}

\end{document}

%% file: data.tex
\section{Hate Speech Datasets}

In this paper, we employ five datasets to assess the capability of LLMs in detecting real-world hate speech. 
Notably, except for CallMeSexist and USElectionHate—whose compositions are influenced by the original size of the datasets—we have generated balanced versions for the remaining selected datasets. Table~\ref{tab:dataset} provides the statistics for the datasets utilized in our experiments.

\noindent \textbf{HateXplain~\cite{mathew2020hatexplain}.} This benchmark hate speech dataset collected datasets from previous hate speech research sources. The data are all collected from Twitter and Gab. The dataset is annotated by Amazon Mechanical Turk (MTurk) workers and contains labels for hate, offensive, and normal categories. Furthermore, it is a multiclass dataset encompassing various types of hate speech, including Race, Religion, Gender, Sexual Orientation, and Miscellaneous. 
In our experiments, we assembled a balanced test dataset by randomly selecting 200 samples from each category of hate speech alongside 1,000 samples labeled as `normal,' resulting in a total of 2,000 test samples. 
Note that all samples labeled `offense' are omitted since we are focusing on the binary classification task.

\noindent \textbf{COVID-HATE~\cite{he2021racism}.} 
The COVID-Hate dataset, gathered from Twitter during the COVID-19 pandemic, specifically focuses on anti-Asian hate speech. It comprises three labels: `anti-Asian,' `counterspeech,' and `neutral.' For our study, we built a test set by randomly selecting 500 samples of anti-Asian hate speech and an equal number of neutral speech samples.

\noindent \textbf{CallMeSexist~\cite{samory2021call}} It was assembled by collecting tweets from Twitter and incorporating data from previous studies. The authors recruited annotators from MTurk to label the tweets as either `sexist' or `other.'  In our work, we utilized all the dataset's 429 samples labeled as `sexist' to represent hate speech, and we randomly added 571 samples from the other category to represent non-hate speech.

\noindent \textbf{USElectionHate~\cite{grimminger-klinger-2021-hate}}
The USElectionHate dataset is culled from Twitter and focuses on dialogues pertinent to the 2020 U.S. Presidential Election and its political discourse. This dataset is intended for binary classification tasks, and in our experiments, we utilize the subset that the original authors have manually annotated. Due to data constraints, we abstained from manually balancing this dataset, unlike our approach with the other datasets under consideration.

\noindent \textbf{SWSR~\cite{JIANG2022100182}} SWSR is the inaugural Chinese dataset focusing on sexism, sourced from Sina Weibo, a social platform similar to Twitter. The dataset has been manually annotated by the authors to categorize comments as either `sexist' or `non-sexist.' To adapt this dataset to our research objectives, we randomly chose 500 comments labeled as `sexist' to represent hate speech and an equal number of `non-sexist' comments to represent non-hate speech.

\begin{table}[t]
    \centering
    \resizebox{0.95\columnwidth}{!}{
    \setlength\tabcolsep{0.65ex} 
    
    \begin{tabular}{ccccc}
    \toprule
        \textbf{Dataset} & \textbf{Size} &\textbf{\begin{tabular}[c]{@{}l@{}}Number of\\hate speech\end{tabular}} & \textbf{Category} &\textbf{Language} \\
        \midrule
         HateXplain~\cite{mathew2020hatexplain}& 2,000 & 1,000 & Multiclass & English \\
         COVID-HATE~\cite{he2021racism}& 1,000 & 500 & Racism & English\\
         CallMeSexist~\cite{samory2021call} &1,000 & 429 & Sexist & English\\
         USElectionHate~\cite{grimminger-klinger-2021-hate}& 600 & 59 & Political & English\\
         SWSR~\cite{JIANG2022100182} & 1,000 & 500 & Sexist & Chinese\\
         \bottomrule
    \end{tabular}
    }
    \caption{Hate speech datasets}
    \label{tab:dataset}
    \vspace{-5mm}
\end{table}

%% file: prompt.tex
\section{Prompt-engineering for Hate Speech Detection}
\label{sec:prompt}

In this section, we elucidate the prompting strategies utilized in our research, offering real-world examples for added clarity, as depicted in Table~\ref{tab:prompts}. 

\noindent \textbf{General prompt.} 
Initially, we use an efficient general prompt previously validated for its effectiveness in hate speech detection~\cite{li2023hot}. Utilizing general prompt engineering techniques allows us to adapt LLMs for the specific task of identifying hate speech. 
To elaborate, for any given sentence $x$, the input to the model under this strategy would be formatted as:
\textit{Do you think this comment is hate speech? comment: \{$x$\} a. Yes b. No}. The model will then output $y$, either ``a. yes'' or ``b. no,'' which represents the classification of the comment as hate speech or non-hate speech, respectively. \looseness=-1

\noindent \textbf{General prompt with hate speech definition.} Furthermore, to account for variations across different datasets, we augment a new general prompt by incorporating a formal definition of hate speech~\cite{10.1371/journal.pone.0221152}, \textit{i.e.}, \textit{Hate speech is speech that attacks a person or group based on attributes such as race, religion, ethnic origin, national origin, sex, disability, sexual orientation, or gender identity}. By providing this additional context, we aim to deepen the LLMs' understanding of what constitutes hate speech and obtain if it can improve their capability in its detection.\looseness=-1

\noindent \textbf{Few-shot learning prompt.} Additionally, with the discussion of the impact of few-shot learning on the performance of LLMs~\cite{10.5555/3495724.3495883}. Many prompts contain example tasks with solutions (few-shot prompting, also referred to as in-context learning). This technique involves supplying the model with a small number of example tasks along with their solutions, essentially `priming' it for the particular type of task at hand. 
We employ ten randomly selected exemplars from the HateXplain dataset as part of the input prompt to assess the efficacy of few-shot prompting for hate speech identification. These ten exemplars are balanced, containing an equal number of hateful and non-hateful samples. We then evaluated the LLM's performance under this setup. One of the exemplars is detailed in Table\ref{tab:prompts} for reference.

\noindent \textbf{Chain-of-thought prompt.} In our study, we also employ Chain-of-Thought (CoT) reasoning, a strategy that interjects intermediate steps---denoted as \(a_1, a_2, \ldots\)---between the input \(x\) and output \(y\). This technique has been previously validated for its effectiveness in enhancing LLMs' complex reasoning abilities~\cite{wei2023chainofthought}. Following a detailed examination of the example prompts in the earlier study, our team of three experts adapted the CoT strategy specifically for detecting hate speech. 
As illustrated in Table~\ref{tab:prompts}, we design our intermediate steps by further elaborating the formal definition of hate speech~\cite{10.1371/journal.pone.0221152} step-by-step, starting from identifying targets to identifying disparaging words. For each example used in the earlier few-shot learning setup, we provide the answer with comprehensive reasoning to clarify if the language should be classified as hateful or not.

\begin{table}[t]
\centering
\resizebox{\columnwidth}{!}{
\setlength\tabcolsep{0.65ex} 
\begin{tabular}{cl}
\toprule
\textbf{Original Sentence} & \blue{You f**king evil pieces of s**t @china} \\
\midrule
\textbf{General prompt} & \begin{tabular}[c]{@{}l@{}}\blue{Do you think this comment is hate speech?} \\ comment: ``You f**king evil pieces of s**t @china"\\ \blue{a. Yes b. No}\end{tabular} \\
\midrule
\textbf{\begin{tabular}[c]{@{}c@{}}General prompt with\\ hate speech definition\end{tabular}} & \begin{tabular}[c]{@{}l@{}}Do you think this comment is hate speech? \\ \blue{The definition of hate speech is `Hate speech' is} \\ \blue{speech that attacks a person or group on the basis}\\ \blue{of attributes such as race, religion, ethnic origin,}\\ \blue{national origin, sex, disability, sexual orientation,}\\ \blue{or gender identity.}\\ comment: ``You f**king evil pieces of s**t @china"\\ a. Yes b. No\end{tabular} \\
\midrule
\textbf{\begin{tabular}[c]{@{}c@{}}Few-shot learning\\ prompt\end{tabular}} & \begin{tabular}[c]{@{}l@{}}\blue{Example 1:} \\ Question: \\ Do you think this comment is hate speech? \\ comment: ``space the final frontier these are the\\ voyages of the starship enterprise its five-year\\ mission to explore strange new worlds to seek out\\ new life and new civilizations to boldly go where\\ no man has gone before where no n***ers are \\ invited nor wanted it a n***er free space"\\ a.Yes b.No \\ \blue{Answer: a.Yes}\\ ... \\ Question:\\ Do you think this comment is hate speech? \\ comment: ``You f**king evil pieces of s**t @china"\\ a. Yes b. No\\\blue{Answer:}\end{tabular} \\
\midrule
\textbf{\begin{tabular}[c]{@{}c@{}}Chain-of-thought \\ prompt \end{tabular}} & \begin{tabular}[c]{@{}l@{}}{\blue{Example 1:}} \\ Question: \\ Do you think this comment is hate speech? \\ comment: ``space the final frontier these are the\\ voyages of the starship enterprise its five-year\\ mission to explore strange new worlds to seek out\\ new life and new civilizations to boldly go where\\ no man has gone before where no n***ers are \\ invited nor wanted it a n***er free space"\\ a.Yes b.No \\ \blue{Answer: This comment is intended to incite hatred}\\ \blue{against the African American community by using}\\ \blue{words such as no n***gers and n***er free to them.}\\ \blue{So the answer is a.Yes}\\ ... \\ Question:\\ Do you think this comment is hate speech? \\ comment: ``You f**king evil pieces of s**t @china"\\ a. Yes b. No\\ \blue{Answer:}\end{tabular}\\
\bottomrule
\end{tabular}%
}
\caption{Prompt samples for four prompting strategies.}
\label{tab:prompts}
\vspace{-5mm}
\end{table}

%% file: evaluation.tex
\section{Measuring the Effectiveness of Prompting Strategies}

\subsection{Experimental Setup}

\noindent \textbf{Baselines.} 
In order to rigorously assess the capability of LLMs in the task of hate speech detection, we have chosen two state-of-the-art baselines for comparative analysis. For the general purpose of hate speech detection, we utilize two transformer models that are fine-tuned on multi-category hate speech datasets.
More specifically, we employ the BERT-base model~\cite{devlin2019bert} that has been fine-tuned explicitly on a well-regarded hate speech classification benchmark dataset, HateXplain\cite{mathew2020hatexplain}, and the fine-tuned variant of RoBERTa model~\cite{liu2019roberta} for another multiclass hate speech dataset, HateCheck~\cite{vidgen2021learning}. 


\noindent \textbf{Model.}
In this study, we employ ChatGPT, specifically the GPT-3.5-turbo variant, as our evaluation model for hate speech detection. Our choice of this model is motivated by several key considerations. 
Firstly, GPT-3.5-turbo stands as one of the most advanced language models that is readily accessible for research, featuring 175 billion parameters that enable it to execute a broad range of text-based tasks with high accuracy. 
Although GPT-4 exists, its limited availability for academic research led us to opt for GPT-3.5-turbo as the most powerful model we could feasibly use. 
Second, this model has exhibited strong performance across multiple natural language processing benchmarks, confirming its generalizability and ability to process and understand context-rich data.
Lastly, GPT-3.5-turbo's pre-trained architecture minimizes the need for task-specific training data, thus reducing the likelihood of introducing bias or errors specific to our training set. We deploy GPT-3.5-turbo across five benchmark datasets to ensure its robustness in capturing the nuances inherent in various forms of hate speech.

\noindent \textbf{Metrics.}
We opt for a comprehensive approach that employs a set of well-known and academically validated metrics to enable a nuanced understanding of model performance. Specifically, we focus on four key evaluation metrics: accuracy, which provides a general measure of the model's correct classifications; precision, which assesses the model's ability to avoid false positives; recall, which evaluates the model's capability to identify true positives; and the F1 score, which offers a balanced measure of the model's precision and recall. Together, these metrics serve as our primary criteria for assessing the performance and effectiveness of the various models under investigation.

\begin{table}[t]
\centering
\resizebox{0.95\columnwidth}{!}{
\setlength\tabcolsep{0.65ex} 
\begin{tabular}{cccccc}
\toprule
\textbf{Model} & \textbf{Dataset} & \textbf{Accuracy} & \textbf{Precision} & \textbf{Recall} & \textbf{F1} \\
 \midrule
\multirow{5}{*}{BERT} & HateXplain & 0.63 & 0.62 & 0.7 & 0.66 \\
& COVID-HATE & 0.51 & 0.46 & \textbf{0.83} & 0.6 \\
 & CallMeSexist & 0.47 & 0.47 & 0.43 & 0.45 \\
 & USElectionHate & 0.33 & 0.1 & \textbf{0.66} & 0.16 \\
 & Total & 0.53 & 0.47 & 0.66 & 0.55 \\
 \midrule
\multirow{5}{*}{RoBERTa} & HateXplain & 0.71 & 0.64 & 0.92 & 0.76 \\
 & COVID-HATE & 0.78 & 0.72 & 0.8 & 0.76 \\
 & CallMeSexist & \textbf{0.7} & 0.76 & \textbf{0.6} & \textbf{0.67} \\
 & USElectionHate & 0.9 & 0.48 & 0.17 & 0.25 \\

 & Total & 0.75 & 0.68 & \textbf{0.79} & \textbf{0.73} \\
  \midrule
\multirow{5}{*}{\begin{tabular}[c]{@{}l@{}}ChatGPT\end{tabular}} & HateXplain & \textbf{0.79} & \textbf{0.72} & \textbf{0.95} & \textbf{0.82} \\
 & COVID-HATE & \textbf{0.81} & \textbf{0.78} & 0.77 & \textbf{0.77} \\
 & CallMeSexist & 0.62 & \textbf{0.81} & 0.31 & 0.44 \\
 & USElectionHate & \textbf{0.91} & \textbf{0.54} & 0.24 & \textbf{0.33} \\
 & Total & \textbf{0.77} & \textbf{0.74} & 0.72 & \textbf{0.73}\\
 \bottomrule
\end{tabular}
}
\caption{Comparison of LLM-based general prompting-strategy with baselines.}
\label{tab:baseline}
\vspace{-5mm}
\end{table}

\subsection{LLM-based General Prompting Strategy vs. Baselines}
In our experiment, we first evaluate the performance of two benchmark models, BERT and RoBERTa, on the hate speech datasets. To explore the capability of LLMs in hate speech detection, we specifically assess ChatGPT, particularly the GPT-3.5-turbo variant, using the general prompt delineated in Section~\ref{sec:prompt}. 
\red{We opt for the general prompting strategy in this evaluation, as it is the most intuitive and fundamental approach typically used by users to interact with LLMs in real-world scenarios.}

As the results depicted in Table~\ref{tab:baseline}, ChatGPT consistently outperforms the fine-tuned BERT model across nearly all metrics. 
For example, in the HateXplain dataset, ChatGPT \red{under the general prompting strategy} exhibits superior accuracy at 0.79 compared to BERT's 0.63. Similarly, ChatGPT's F1 Score of 0.82 surpasses BERT's 0.66. This trend generally continues across other datasets, leading to higher overall totals for ChatGPT relative to BERT.
Compared to the fine-tuned RoBERTa model, ChatGPT demonstrates competitive or superior performance in most metrics. On the HateXplain dataset, it outperforms RoBERTa with an accuracy of 0.79 against 0.71 and an F1 Score of 0.82 against 0.76. While RoBERTa closely matches ChatGPT in the total F1 Score of 0.73, ChatGPT edges it out with an F1 Score of 0.77.
While ChatGPT consistently outperforms the fine-tuned BERT model and demonstrates competitive or superior performance to the fine-tuned RoBERTa across multiple datasets, it's important to note some limitations. 
\red{Specifically, ChatGPT with the general prompt shows lower recall scores in datasets like USElectionHate and CallMeSexist, suggesting that the model prompted with just a general prompt may not be able to capture sufficient contextual knowledge of specific types of hate speech. 
Despite these limitations, the remarkable accuracy and F1 scores across most datasets suggest that it is generally adequate for online hate speech detection, even if there are specific areas where further research could improve its performance.}

\subsection{Analysis of Different Prompts}
To evaluate the impact of various prompt engineering techniques on the effectiveness of Large Language Models (LLMs), we employ the HateXplain benchmark dataset, as detailed in Table~\ref{tab:prompts}, to assess ChatGPT's performance. 
We concentrate on this dataset because it is a multiclass repository encompassing all types of hate speech present in other datasets.
Moreover, using the benchmark dataset allows us to have one standardized criterion for annotating different categories of hate speech.

In our experiment, we utilize four distinct prompts: the General Prompt (GP), General Prompt with Hate Speech Definition (GPwDef), Few-Shot Learning Prompt (Few-Shot), and Chain-of-Thought Reasoning Prompt (CoT).
As evidenced by the results in Table~\ref{tab:comparison}, different prompting strategies exhibit varying levels of effectiveness in guiding ChatGPT for hate speech detection. CoT outperforms the other prompts across all metrics—achieving an accuracy of 0.85, precision of 0.8, recall of 0.95, and an F1 score of 0.87. This suggests that providing a structured chain of reasoning within the prompt substantially enhances the model's understanding and detection of hate speech. \red{Furthermore, breaking up the problem of hate speech detection into multiple steps could better leverage the model's learned knowledge base, thus effectively using it to make contextual decisions.}
In contrast, the Few-Shot Learning Prompt (Few-Shot) performs the least effectively among the four prompts, particularly in terms of accuracy (0.74) and precision (0.67). Despite this, it scores fairly well on recall (0.97), indicating that it is sensitive but less precise in identifying hate speech instances. \red{We posit that the suboptimal performance may be attributable to the limited number of randomly selected exemplars used in the prompt, which may not adequately capture the full context of the diverse types of hate speech.} Addressing this limitation could potentially enhance the model's precision and overall effectiveness performance.
Another interesting finding is that the General Prompt (GP) and General Prompt with Hate Speech Definition (GPwDef) offer comparable performances, with GPwDef slightly lagging behind GP. Although incorporating a formal definition into the prompt (GPwDef) was hypothesized to augment the model's understanding of hate speech, the impact appears to be minimal based on our metrics.

\begin{table}[b]
\vspace{-3mm}
\centering
\resizebox{0.95\columnwidth}{!}{
\begin{tabular}{ccccc}
\toprule
\textbf{Prompt} & \textbf{Accuracy} & \textbf{Precision} & \textbf{Recall} & \textbf{F1} \\
\midrule
GP              & 0.79              & 0.72               & 0.95            & 0.82        \\
GPwDef          & 0.78              & 0.71               & 0.95            & 0.81        \\
Few-Shot        & 0.74              & 0.67               & \textbf{0.97}            & 0.79         \\
CoT             & \textbf{0.85 }             & \textbf{0.8 }               & 0.95            & \textbf{0.87}       \\
\bottomrule
\end{tabular}
}
\caption{Comparison of different prompting strategies.}
\label{tab:comparison}
\end{table}



This empirical analysis underscores the critical importance of prompt engineering in tuning LLMs for specialized tasks like hate speech detection. Specifically, carefully designing a prompt that fully captures the complexity and nuance of the task can significantly improve the model's performance.

\subsection{Effectiveness of LLMs against multilingual hate speech}
In light of LLMs' exceptional performance in multilingual machine translation tasks~\cite{zhu2023multilingual}, our study also explores the models' ability to recognize hate speech in languages other than English. 
For this purpose, we employ the sexism-related Chinese hate speech dataset SWSR to evaluate ChatGPT. 
As illustrated in Table~\ref{tab:chinese}, the four prompts we tested exhibit comparable accuracy levels, ranging from 0.61 to 0.68. 
While these scores indicate a basic proficiency in understanding Chinese text, the low Recall and F1 scores reveal that LLMs face challenges when it comes to hate speech detection in Chinese languages. 
Most notably, CoT, which excels in English hate speech detection, performs poorly on Chinese hate speech. 
Although CoT impressively maintains a high Precision score of 0.86, suggesting that when it does engage with and classify sentences, it does so with high accuracy, this is counterbalanced by its low Recall and F1 scores of 0.21 and 0.34, indicating that the model frequently fails to identify instances of hate speech that should be flagged. 
This discrepancy in performance could be due to the underlying linguistic and cultural factors that CoT might not capture effectively in a non-English setting. 
Another potential explanation for this low performance could be that the intermediate reasoning steps, being in English, introduce an additional layer of complexity that confounds the model's understanding of multilingual content. 
This divergence highlights an area of concern: while LLMs can understand multiple languages, their proficiency in specialized tasks like hate speech detection can vary significantly across languages.


\begin{table}[t]
\centering
\resizebox{0.95\columnwidth}{!}{
\begin{tabular}{ccccc}
\toprule
\textbf{Prompt}  & \textbf{Accuracy} & \textbf{Precision} & \textbf{Recall} & \textbf{F1} \\
\midrule
GP                              & \textbf{0.68 }             & 0.73               & 0.36            & 0.48        \\
GPwDef                             & 0.61                  & 0.78                    & 0.31                & 0.44             \\
Few-Shot                         & 0.66              & 0.81               & \textbf{0.42}            & \textbf{0.55}          \\
CoT                           & 0.64              & \textbf{0.86}               & 0.21            & 0.34     \\
\bottomrule
\end{tabular}
}
\caption{LLMs against Chinese language hate speech.}
\label{tab:chinese}
\vspace{-5mm}
\end{table}

%% file: main.bbl
\begin{thebibliography}{10}
\providecommand{\url}[1]{#1}
\csname url@samestyle\endcsname
\providecommand{\newblock}{\relax}
\providecommand{\bibinfo}[2]{#2}
\providecommand{\BIBentrySTDinterwordspacing}{\spaceskip=0pt\relax}
\providecommand{\BIBentryALTinterwordstretchfactor}{4}
\providecommand{\BIBentryALTinterwordspacing}{\spaceskip=\fontdimen2\font plus
\BIBentryALTinterwordstretchfactor\fontdimen3\font minus \fontdimen4\font\relax}
\providecommand{\BIBforeignlanguage}[2]{{%
\expandafter\ifx\csname l@#1\endcsname\relax
\typeout{** WARNING: IEEEtran.bst: No hyphenation pattern has been}%
\typeout{** loaded for the language `#1'. Using the pattern for}%
\typeout{** the default language instead.}%
\else
\language=\csname l@#1\endcsname
\fi
#2}}
\providecommand{\BIBdecl}{\relax}
\BIBdecl

\bibitem{thomas2021sok}
K.~Thomas, D.~Akhawe, M.~Bailey, D.~Boneh, E.~Bursztein, S.~Consolvo, N.~Dell, Z.~Durumeric, P.~G. Kelley, D.~Kumar \emph{et~al.}, ``Sok: Hate, harassment, and the changing landscape of online abuse,'' in \emph{2021 IEEE Symposium on Security and Privacy (SP)}.\hskip 1em plus 0.5em minus 0.4em\relax IEEE, 2021, pp. 247--267.

\bibitem{10.1371/journal.pone.0221152}
\BIBentryALTinterwordspacing
S.~MacAvaney, H.-R. Yao, E.~Yang, K.~Russell, N.~Goharian, and O.~Frieder, ``Hate speech detection: Challenges and solutions,'' \emph{PLOS ONE}, vol.~14, no.~8, pp. 1--16, 08 2019. [Online]. Available: \url{https://doi.org/10.1371/journal.pone.0221152}
\BIBentrySTDinterwordspacing

\bibitem{twitter_hate_speech_policy}
\BIBentryALTinterwordspacing
``Hateful conduct policy,'' accessed: 2022-08-06. [Online]. Available: \url{https://help.twitter.com/en/rules-and-policies/hateful-conduct-policy}
\BIBentrySTDinterwordspacing

\bibitem{twitter_hate_speech_AI}
\BIBentryALTinterwordspacing
``Twitter says ai flags over half of tweets violating terms of services,'' accessed: 2022-08-06. [Online]. Available: \url{https://www.emergingtechbrew.com/stories/2020/07/20/twitter-says-ai-flags-half-tweets-violating-terms-services}
\BIBentrySTDinterwordspacing

\bibitem{facebook_hate_speech_AI}
\BIBentryALTinterwordspacing
``How facebook uses super-efficient ai models to detect hate speech,'' accessed: 2022-08-06. [Online]. Available: \url{https://ai.facebook.com/blog/how-facebook-uses-super-efficient-ai-models-to-detect-hate-speech/}
\BIBentrySTDinterwordspacing

\bibitem{elsherief2018hate}
M.~ElSherief, V.~Kulkarni, D.~Nguyen, W.~Y. Wang, and E.~Belding, ``Hate lingo: A target-based linguistic analysis of hate speech in social media,'' in \emph{Proceedings of the International AAAI Conference on Web and Social Media}, vol.~12, no.~1, 2018.

\bibitem{nobata2016abusive}
C.~Nobata, J.~Tetreault, A.~Thomas, Y.~Mehdad, and Y.~Chang, ``Abusive language detection in online user content,'' in \emph{Proceedings of the 25th international conference on world wide web}, 2016, pp. 145--153.

\bibitem{schmidt-wiegand-2017-survey}
\BIBentryALTinterwordspacing
A.~Schmidt and M.~Wiegand, ``A survey on hate speech detection using natural language processing,'' in \emph{Proceedings of the Fifth International Workshop on Natural Language Processing for Social Media}.\hskip 1em plus 0.5em minus 0.4em\relax Valencia, Spain: Association for Computational Linguistics, Apr. 2017, pp. 1--10. [Online]. Available: \url{https://aclanthology.org/W17-1101}
\BIBentrySTDinterwordspacing

\bibitem{openai2022gptblog}
OpenAI, ``Introducing chatgpt,'' \url{https://openai.com/blog/chatgpt/}, 2022.

\bibitem{bang2023multitask}
Y.~Bang, S.~Cahyawijaya, N.~Lee, W.~Dai, D.~Su, B.~Wilie, H.~Lovenia, Z.~Ji, T.~Yu, W.~Chung, Q.~V. Do, Y.~Xu, and P.~Fung, ``A multitask, multilingual, multimodal evaluation of chatgpt on reasoning, hallucination, and interactivity,'' 2023.

\bibitem{guo2023close}
B.~Guo, X.~Zhang, Z.~Wang, M.~Jiang, J.~Nie, Y.~Ding, J.~Yue, and Y.~Wu, ``How close is chatgpt to human experts? comparison corpus, evaluation, and detection,'' 2023.

\bibitem{zhao2023survey}
W.~X. Zhao, K.~Zhou, J.~Li, T.~Tang, X.~Wang, Y.~Hou, Y.~Min, B.~Zhang, J.~Zhang, Z.~Dong, Y.~Du, C.~Yang, Y.~Chen, Z.~Chen, J.~Jiang, R.~Ren, Y.~Li, X.~Tang, Z.~Liu, P.~Liu, J.-Y. Nie, and J.-R. Wen, ``A survey of large language models,'' 2023.

\bibitem{10.1145/3560815}
\BIBentryALTinterwordspacing
P.~Liu, W.~Yuan, J.~Fu, Z.~Jiang, H.~Hayashi, and G.~Neubig, ``Pre-train, prompt, and predict: A systematic survey of prompting methods in natural language processing,'' \emph{ACM Comput. Surv.}, vol.~55, no.~9, jan 2023. [Online]. Available: \url{https://doi.org/10.1145/3560815}
\BIBentrySTDinterwordspacing

\bibitem{doj_hate_crime_laws}
``Hate crime laws,'' \url{https://www.justice.gov/crt/hate-crime-laws}, accessed: 2022-08-14.

\bibitem{fb_hate_speech_def}
\BIBentryALTinterwordspacing
``{Hate Speech},'' 2021, accessed: 2022-08-03. [Online]. Available: \url{https://transparency.fb.com/policies/community-standards/hate-speech/?from=https\%3A\%2F\%2Fm.facebook.com\%2Fcommunitystandards\%2Fhate_speech\%2F&refsrc=deprecated}
\BIBentrySTDinterwordspacing

\bibitem{duggan2017online}
M.~Duggan, ``Online harassment 2017,'' 2017.

\bibitem{microsoft_civility}
\BIBentryALTinterwordspacing
``Microsoft, ``civility, safety \& interaction online'','' accessed: 2022-03-03. [Online]. Available: \url{https://news.microsoft.com/wp-content/uploads/prod/sites/421/2020/02/Digital-Civility-2020-Global-Report.pdf}
\BIBentrySTDinterwordspacing

\bibitem{fb_community_standards_enforcement_report}
``Facebook community standards enforcement report,'' \url{https://transparency.fb.com/data/community-standards-enforcement/}, note = {Accessed: 2022-08-14}.

\bibitem{healthier_twitter}
``A healthier twitter: Progress and more to do,'' \url{https://blog.twitter.com/en_us/topics/company/2019/health-update}, accessed: 2022-08-14.

\bibitem{content_moderators}
\BIBentryALTinterwordspacing
``{Content moderators at YouTube, Facebook and Twitter see the worst of the web and suffer silently},'' accessed: 2022-08-13. [Online]. Available: \url{{https://www.washingtonpost.com/technology/2019/07/25/social-media-companies-are-outsourcing-their-dirty-work-philippines-generation-workers-is-paying-price/}}
\BIBentrySTDinterwordspacing

\bibitem{chiu2022detecting}
K.-L. Chiu, A.~Collins, and R.~Alexander, ``Detecting hate speech with gpt-3,'' 2022.

\bibitem{li2023hot}
L.~Li, L.~Fan, S.~Atreja, and L.~Hemphill, ``"hot" chatgpt: The promise of chatgpt in detecting and discriminating hateful, offensive, and toxic comments on social media,'' 2023.

\bibitem{he2023prompt}
X.~He, S.~Zannettou, Y.~Shen, and Y.~Zhang, ``You only prompt once: On the capabilities of prompt learning on large language models to tackle toxic content,'' 2023.

\bibitem{zhong2023chatgpt}
Q.~Zhong, L.~Ding, J.~Liu, B.~Du, and D.~Tao, ``Can chatgpt understand too? a comparative study on chatgpt and fine-tuned bert,'' 2023.

\bibitem{Huang_2023}
\BIBentryALTinterwordspacing
F.~Huang, H.~Kwak, and J.~An, ``Chain of explanation: New prompting method to generate quality natural language explanation for implicit hate speech,'' in \emph{Companion Proceedings of the {ACM} Web Conference 2023}.\hskip 1em plus 0.5em minus 0.4em\relax {ACM}, apr 2023. [Online]. Available: \url{https://doi.org/10.1145%2F3543873.3587320}
\BIBentrySTDinterwordspacing

\bibitem{mathew2020hatexplain}
B.~Mathew, P.~Saha, S.~M. Yimam, C.~Biemann, P.~Goyal, and A.~Mukherjee, ``Hatexplain: A benchmark dataset for explainable hate speech detection,'' \emph{arXiv preprint arXiv:2012.10289}, 2020.

\bibitem{he2021racism}
B.~He, C.~Ziems, S.~Soni, N.~Ramakrishnan, D.~Yang, and S.~Kumar, ``Racism is a virus: Anti-asian hate and counterspeech in social media during the covid-19 crisis,'' 2021.

\bibitem{samory2021call}
M.~Samory, I.~Sen, J.~Kohne, F.~Floeck, and C.~Wagner, ````call me sexist, but...": Revisiting sexism detection using psychological scales and adversarial samples,'' 2021.

\bibitem{grimminger-klinger-2021-hate}
\BIBentryALTinterwordspacing
L.~Grimminger and R.~Klinger, ``Hate towards the political opponent: A {T}witter corpus study of the 2020 {US} elections on the basis of offensive speech and stance detection,'' in \emph{Proceedings of the Eleventh Workshop on Computational Approaches to Subjectivity, Sentiment and Social Media Analysis}.\hskip 1em plus 0.5em minus 0.4em\relax Online: Association for Computational Linguistics, Apr. 2021, pp. 171--180. [Online]. Available: \url{"https://aclanthology.org/2021.wassa-1.18"}
\BIBentrySTDinterwordspacing

\bibitem{JIANG2022100182}
\BIBentryALTinterwordspacing
A.~Jiang, X.~Yang, Y.~Liu, and A.~Zubiaga, ``Swsr: A chinese dataset and lexicon for online sexism detection,'' \emph{Online Social Networks and Media}, vol.~27, p. 100182, 2022. [Online]. Available: \url{https://www.sciencedirect.com/science/article/pii/S2468696421000604}
\BIBentrySTDinterwordspacing

\bibitem{10.5555/3495724.3495883}
T.~B. Brown, B.~Mann, N.~Ryder, M.~Subbiah, J.~Kaplan, P.~Dhariwal, A.~Neelakantan, P.~Shyam, G.~Sastry, A.~Askell, S.~Agarwal, A.~Herbert-Voss, G.~Krueger, T.~Henighan, R.~Child, A.~Ramesh, D.~M. Ziegler, J.~Wu, C.~Winter, C.~Hesse, M.~Chen, E.~Sigler, M.~Litwin, S.~Gray, B.~Chess, J.~Clark, C.~Berner, S.~McCandlish, A.~Radford, I.~Sutskever, and D.~Amodei, ``Language models are few-shot learners,'' in \emph{Proceedings of the 34th International Conference on Neural Information Processing Systems}, ser. NIPS'20.\hskip 1em plus 0.5em minus 0.4em\relax Red Hook, NY, USA: Curran Associates Inc., 2020.

\bibitem{wei2023chainofthought}
J.~Wei, X.~Wang, D.~Schuurmans, M.~Bosma, B.~Ichter, F.~Xia, E.~Chi, Q.~Le, and D.~Zhou, ``Chain-of-thought prompting elicits reasoning in large language models,'' 2023.

\bibitem{devlin2019bert}
J.~Devlin, M.-W. Chang, K.~Lee, and K.~Toutanova, ``Bert: Pre-training of deep bidirectional transformers for language understanding,'' 2019.

\bibitem{liu2019roberta}
Y.~Liu, M.~Ott, N.~Goyal, J.~Du, M.~Joshi, D.~Chen, O.~Levy, M.~Lewis, L.~Zettlemoyer, and V.~Stoyanov, ``Roberta: A robustly optimized bert pretraining approach,'' 2019.

\bibitem{vidgen2021learning}
B.~Vidgen, T.~Thrush, Z.~Waseem, and D.~Kiela, ``Learning from the worst: Dynamically generated datasets to improve online hate detection,'' 2021.

\bibitem{zhu2023multilingual}
W.~Zhu, H.~Liu, Q.~Dong, J.~Xu, S.~Huang, L.~Kong, J.~Chen, and L.~Li, ``Multilingual machine translation with large language models: Empirical results and analysis,'' 2023.

\bibitem{10179315}
Y.~Qu, X.~He, S.~Pierson, M.~Backes, Y.~Zhang, and S.~Zannettou, ``On the evolution of (hateful) memes by means of multimodal contrastive learning,'' in \emph{2023 IEEE Symposium on Security and Privacy (SP)}, 2023, pp. 293--310.

\bibitem{kiela2021hateful}
D.~Kiela, H.~Firooz, A.~Mohan, V.~Goswami, A.~Singh, P.~Ringshia, and D.~Testuggine, ``The hateful memes challenge: Detecting hate speech in multimodal memes,'' 2021.

\bibitem{osti_10399964}
\BIBentryALTinterwordspacing
K.~Cuo, W.~Zhao, M.~Jaden, V.~Vishwamitra, Z.~Zhao, and H.~Hu, ``Understanding the generalizability of hateful memes detection models against covid-19-related hateful memes,'' \emph{International Conference on Machine Learning and Applications}. [Online]. Available: \url{https://par.nsf.gov/biblio/10399964}
\BIBentrySTDinterwordspacing

\end{thebibliography}
